\documentclass[fleqn,12pt,twoside]{article}

\usepackage{espcrc1}
\usepackage{graphicx}

\newcommand{\AmS}{{\protect\the\textfont2
  A\kern-.1667em\lower.5ex\hbox{M}\kern-.125emS}}

\title{Dynamical coupled-channel study of $K^+ \Lambda$ photoproduction}

\author{B. Juli\'a-D\'{\i}az\address[MCSD]{Department of Physics and Astronomy,
University of Pittsburgh, PA 15260, USA}\thanks{National Science Foundation,
grant No. 0244526 at the University of Pittsburgh.},
        B. Saghai\address{D\'epartement d'Astrophysique, de Physique des Particules,
de Physique Nucl\'eaire et de\\
l'Instrumentation Associ\'ee, DSM, CEA/Saclay, 91191 Gif-sur-Yvette,
France}\thanks{Thanks the Organizer for the oral presentation opportunity of the
         present work during the Conference.},
        F. Tabakin\addressmark[MCSD]\thanks{National Science Foundation,
grant No. 0244526 at the University of Pittsburgh.},
        W.-T. Chiang\address{Central Research and Development Division, 
        UMC, Hsinchu 30077, Taiwan}
        T.-S. H. Lee\address{Physics Division, Argonne National Laboratory, Argonne,
               IL 60439, USA}\thanks{U.S. Department of Energy, Office of Nuclear Physics, 
               Contract No. W-31-109-ENG-38.}
        and
        Z. Li\address{Department of Computer and Information Science, University of 
        Maryland, MD 20783, USA}
        }

\begin{document}

\maketitle

\begin{abstract}
Results for the reaction $ \gamma p \rightarrow K^+  \Lambda$,
studied within a constituent quark model and a dynamical
coupled-channel approach, are presented and compared with recent
data. Issues related to the search for missing baryon resonances
are briefly discussed and the role played by a third $S_{11}$
resonance is underlined.
\end{abstract}

\section{INTRODUCTION}

The electromagnetic production of associated strangeness on the
proton in the total center-of-mass range from threshold up to
$W\approx$~2.3 GeV is under extensive study both experimentally
and theoretically. Besides understanding the elementary reaction
mechanism, a strong motivation is the search for the missing
baryon resonances and improved insights into the known ones.

The quality of recent data~\cite{JLab1,JLab2,ELSA,LEPS,Graal}
requires going beyond the direct production channel, by taking
into account intermediate and final state interactions
(FSI)~\cite{CC-01,CC-04,CC}. In an earlier paper~\cite{CC-01}, it
was reported that coupled-channel~(CC) effects are significant at
the level of inducing up to 20\% changes on total cross sections.
In this latter investigation the direct (without CC effects)
channel $ \gamma p \rightarrow K^+  \Lambda$ was described via a
very simple effective Lagrangian model, which embodied only two
baryon resonances. The present work takes advantage of a
comprehensive chiral constituent model (CQM)~\cite{CQM}, which
includes all known baryon resonances. Moreover, for the CC
effects, here we use a more advanced model~\cite{CC-04} for
meson-baryon interactions to take into account the multi-step
processes such as $\gamma N \to \pi N \to KY$ and $\gamma N \to KY
\to KY$.

\section{THEORETICAL FRAME }

Here we recall very briefly how the direct $K^+ \Lambda$
production, via a CQM, and the CC approach, accounts for
strangeness production including $KY\to KY$ FSI, as well as $\pi N
\to KY$ and $KY\to KY$ intermediate state interactions (ISI); with
$Y \equiv \Lambda, \Sigma$.
\subsection{Direct channel: chiral constituent quark formalism}
The low energy QCD Lagrangian is the starting point of this
approach to pseudoscalar meson photoproduction on nucleons, based
on the $SU(6)\otimes O(3)$ symmetry. The present work goes beyond
the exact $SU(6)\otimes O(3)$ symmetry by introducing~\cite{CQM}
the configuration mixing generated by gluon exchange
interactions~\cite{IK}.

The advantage of the quark model is the ability to relate the
photoproduction data directly to the internal structure of the
baryon resonances. It also allows one to introduce in the reaction
mechanism {\it all} known nucleon ({\it s-}channel) and hyperon
({\it u-}channel) resonances.

This formalism is proven~\cite{CQM} to produce realistic direct
channel models in good agreement with the data for both $ \gamma p
\to \eta p$ and $ \gamma p \to K^+  \Lambda$ processes and shows
the need for a third $S_{11}$ resonance.
%
%
\subsection{Coupled-channel intermediate and final state meson-baryon interactions}
The intermediate state reactions ($\pi N \rightarrow KY$ and $KY \rightarrow KY$)
are studied~\cite{CC-04} using a dynamical coupled-channel model of meson-baryon
interactions at energies where the baryon resonances are strongly excited.
The channels included are: $\pi N$, $K\Lambda$, and $K\Sigma$.
The resonances considered are:
$N^*$ [$S_{11}(1650)$, $P_{11}(1710)$, $P_{13}(1720)$,
$D_{13}(1700)$]; $\Delta^*$ [$S_{31}(1900)$, $P_{31}(1910)$,
$P_{33}(1920)$]; $\Lambda ^*$ [$S_{01}(1670)$, $P_{01}(1810)$];
$\Sigma^*$ [$P_{11}(1660)$, $D_{13}(1670)$]; and $K^*$(892).

The basic non-resonant $\pi N \rightarrow KY$ and $KY\rightarrow KY$ transition
potentials are derived from effective Lagrangians using a unitary transformation
method. The dynamical coupled-channel equations are simplified by parametrizing
the $\gamma N \to \pi N$ and $\pi N \to \pi N$ amplitudes in terms of empirical
$\pi N$ partial-wave amplitudes~\cite{SAID} and a phenomenological off-shell function.
A model has been constructed
with the coupling constants and resonance parameters consistent with the SU(3)
symmetry and/or the Particle Data Group values~\cite{PDG}.
Good fits to the available data for
$\pi^- p \to K^\circ \Lambda,~K^\circ \Sigma^\circ$
have been achieved~\cite{CC-04}.
%
%
\section{RESULTS AND DISCUSSION}
We focus on the interpretation of recent data from the CLAS
Collaboration~\cite{JLab1}, embodying all 920 measured
differential cross sections~\cite{JLab2} in our data-base.
This latter has been fitted, using the CERN MINUIT package. The
free parameters are those of the CQM (mainly one $SU(6)\otimes O(3)$
symmetry breaking parameter per nucleon resonance with $M \leq$ 2 GeV).
Parameters for the intermediate and final state meson-baryon interactions
are taken from the model B in Ref.~\cite{CC-04}.

\begin{figure}[t!]
\includegraphics[width=160mm]{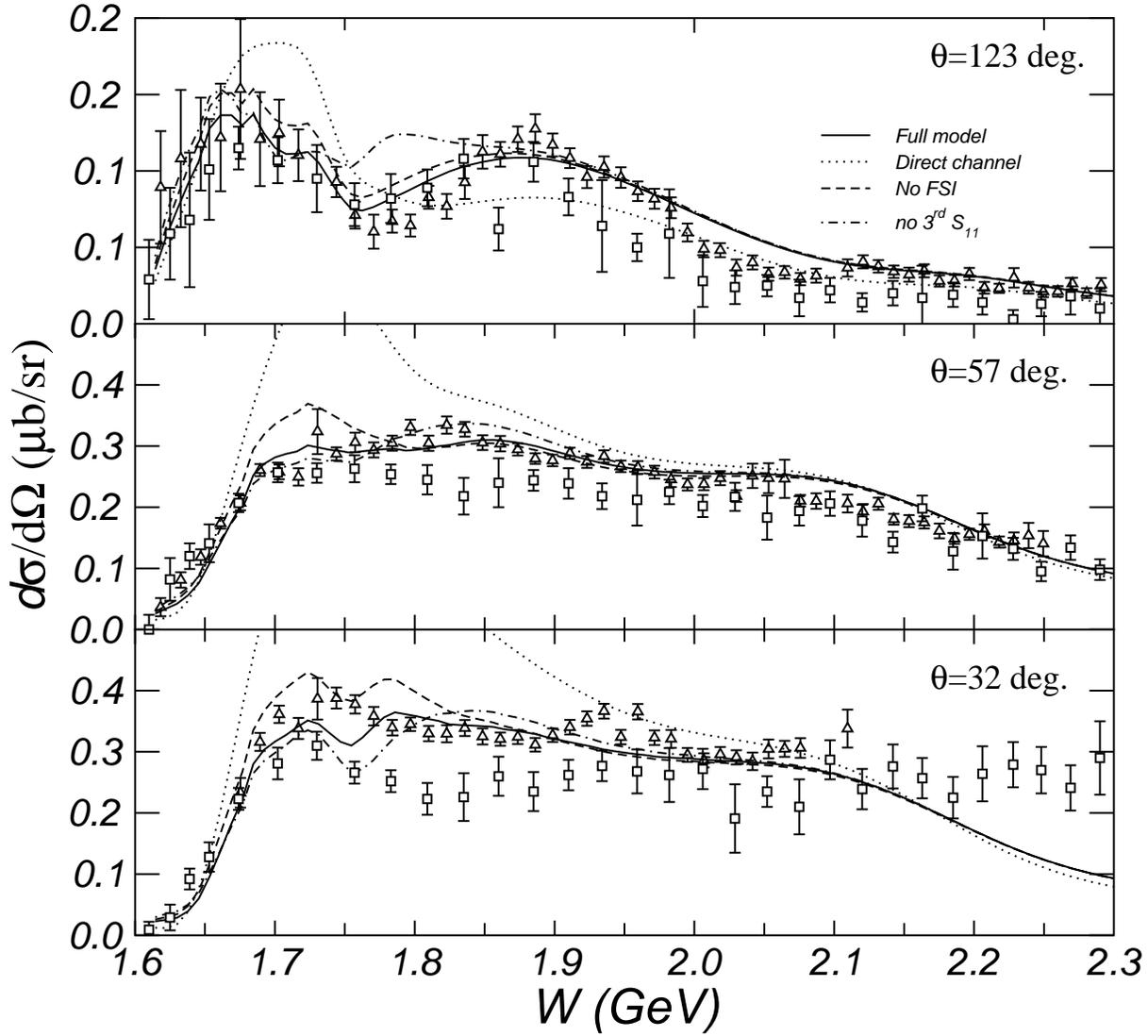}
\caption{Excitation functions as a function of total center-of-mass energy at
$\theta =32^\circ$, $57^\circ$ and $123^\circ$. Full curves: complete results
(CQM+CC).
Dotted curves: direct channel (without CC effects).
Dashed curves: full calculation, but without final state $KY \to KY$ interactions.
Dot-dashed curves: full calculation, but without the third $S_{11}$resonance.
Fitted data (Triangles) are from CLAS ~\protect~\cite{JLab1}. SAPHIR
data~\protect~\cite{ELSA} are also depicted (Squares).}
\label{fig:largenenough}
\end{figure}

Results for the differential cross section at three
angles are shown in Figure~\ref{fig:largenenough}.
The full curves come from our complete model,
embodying all relevant ISI and FSI,
and reproduce the 920 data points quite well.
In order to emphasize different effects in the reaction under study,
three other curves are depicted. They are obtained without further
minimizations, but only by swicthing off some of the parameters related to
specific contributions.

The dotted curves (Direct channel), are obtained by switching off the coupled-channel
interactions, keeping hence only the direct channel contributions.
The sizeable effects observed in the second resonance region show clearly
the crucial role played by the CC phenomena.
%
%

The dashed curves (No FSI) come from the CC effects only due to the intermediate
$\pi N \to \pi N,~KY$ interactions. Here final state $KY \to KY$ are
discarded.
The results show very significant contributions from the ISI.
As expected, the FSI are less important,
but they are not negligible.

Going back to the full model, we have switched off the third
$S_{11}$ resonance, see the dot-dashed curves. Significant
contributions, especially at the most backward angle, are observed
around 1.8 GeV. The extracted values for the mass and the width of
this resonance, via the minimization procedure, are $M
\approx$~1.780 GeV and $\Gamma \approx$~100 MeV, respectively.
Those values agree with our previous findings~\cite{CQM}, as well
as other results reported~\cite{S11} in the literature.

We have also investigated possible contributions from a third $P_{13}$ missing
resonance, but found no significant effects.

Concerning the data-base, the significant discrepencies between
the CLAS~\cite{JLab1,JLab2} and SAPHIR~\cite{ELSA} Collaborations
does not allow a simultaneous fitting of both data-sets. However,
fitting seperately the SAPHIR data, leads to slightly different
free parameters, but to comparable conclusions on the effects
investigated via the curves shown in Fig. 1. The existing and
forthcoming polarization data~\cite{JLab1,JLab2,ELSA,LEPS,Graal} will, hopefully,
clear up those experimental issues.

In summary, we have shown, on the one hand, the crucial need for coupled-channel
approaches in the electromagnetic production of mesons, and on the other hand,
confirmed significant contributions from a third $S_{11}$ nucleon resonance to the
reaction mechanism.
%

\end{document}